\documentclass{ws-procs9x6}
\def\lsim{\mathrel{\rlap{\lower4pt\hbox{\hskip1pt$\sim$}}
    \raise1pt\hbox{$<$}}}         
\def\gsim{\mathrel{\rlap{\lower4pt\hbox{\hskip1pt$\sim$}}
    \raise1pt\hbox{$>$}}}         

\begin{document}

\title{HARMONIC-OSCILLATOR-BASED EFFECTIVE THEORY}

\author{W. C. HAXTON}

\address{Inst. for Nuclear Theory and Dept. of Physics,
University of Washington\\
Seattle, WA 98195, USA\\
E-mail: haxton@phys.washington.edu}

\begin{abstract}
I describe harmonic-oscillator-based effective theory (HOBET) and explore the
extent to which the effects of excluded higher-energy oscillator shells can be
represented by a contact-gradient expansion in next-to-next-to-leading order
(NNLO).   I find the expansion can be very successful provided the 
energy dependence of the effective interaction, connected with missing 
long-wavelength physics associated with low-energy breakup channels, 
is taken into account.   I discuss a
modification that removes operator mixing from HOBET, simplifying the task
of determining the parameters of an NNLO interaction.
\end{abstract}


\bodymatter

\section{Introduction}
Often the problem of calculating long-wavelength
nuclear observables -- binding energies, radii, or responses to low-momentum
probes -- is formulated in terms of pointlike, nonrelativistic nucleons 
interacting through a potential.  To solve this problem theorists have developed both
nuclear models, which are not
systematically improvable, and exact numerical techniques, such as
fermion Monte Carlo.  Because the nuclear many-body problem
is so difficult -- one must simultaneously deal with anomalously large
NN scattering lengths and a potential that has a short-range, strongly repulsive core --
exact approaches are numerically challenging, so far limited to the lighter nuclei within
the 1s and lower 1p shells.  The Argonne theory group has been one of the main
developers of such exact methods \cite{argonne}.

However, effective theory (ET) offers an alternative, a method that limits the numerical difficulty
of a calculation by restricting it to a finite Hilbert space (the $P$- or
``included"-space), while correcting the bare Hamiltonian $H$ (and other operators) for the
effects of the $Q$- or ``excluded"-space. 
Calculations using the effective Hamiltonian $H^{eff}$ within $P$ reproduce
the results using $H$ within $P+Q$, over the domain of overlap. 
That is, the effects of $Q$ on $P$-space 
calculations are absorbed into $P(H^{eff}-H)P$.   There may
exist some systematic expansion -- perhaps having to do with the shorter range of
interactions in $Q$ -- that simplifies the determination $P(H^{eff}-H)P$ \cite{weinberg,savage}.

One interesting challenge for ET is the case of a $P$-space basis of harmonic
oscillator (HO) Slater determinants.  This is a special basis for nuclear physics
because of center-of-mass separability:  if all Slater determinants containing
up to $n$ oscillator quanta are retained, $H^{eff}$ will be translationally invariant (assuming
$H$ is).  Such bases are also important because of powerful shell-model (SM) techniques that 
have been developed for interative diagonalization and for evaluating inclusive responses.  The larger
$P$ can be made, the smaller the effects of $H^{eff}-H$. 

There are two common approaches to the ET problem.  One is 
the determination of $P(H^{eff}-H)P$ from a given $H$ known throughout $P+Q$, a 
problem that appears naively to be no less difficult than the original $P+Q$ diagonalization of $H$.  
However, this may not be the case if $H$ is somehow simpler when acting in $Q$.  
For example, if $Q$ contains primarily
high-momentum (short-distance) interactions, then $H^{eff}-H$ might have a cluster
expansion:  it becomes increasingly unlikely to have $m$ nucleons in close proximity,
as $m$ increases (e.g., a maximum of four nucleons can be in a relative s-state).  Thus one could
approximate the full scattering series in $Q$ by successive two-body, three-body,
etc., terms, with the
expectation that this series will converge quickly with increasing nucleon number.
This would explain why simple two-nucleon ladder sums -- 
g-matrices -- have been 
somewhat successful as effective interactions \cite{kuobrown} (however, see
Ref.~\refcite{barrett,shucan}).

The second approach is that usually taken in effective field theories \cite{savage},
determining $H^{eff}$ phenomenologically.
This is the ``eliminate the middleman" approach: 
$H$ itself is an effective interaction, parameterized in
order to reproduce $NN$ scattering and other data up to some energy.  So why go to the
extra work of this intermediate stage between QCD and SM-like spaces?  
This alternative approach begins with $PHP$, the long-range $NN$ interaction that is 
dominated by pion exchange and constrained by chiral symmetry. 
The effects of the omitted $Q$-space, $P(H^{eff}-H)P$, might be expressed in some
systematic expansion, with the coefficients of that expansion directly determined
from data, rather than from any knowledge of $H$ acting outside of $P$.

While we explored this second approach some years ago, some subtle issues arose,
connected with properties of HO bases.   This convinced us that the first step in our
program should be solving and thoroughly understanding the effective
interactions problem via the first approach, so that we would have answers in hand 
to test the success of more phenomenological approaches.  Thus we proceeded 
to follow the first approach using a realistic $NN$ potential, $av18$ \cite{argonne2},
generating $H^{eff}$ numerically for the two- and three-body problems in
a variety of HO SM spaces.  Here I will use these results to show that a
properly defined
short-range interaction provides an excellent representation of the
effective interaction.  This is an encouraging result, one that suggests 
a purely phenomenological treatment of the effective interaction might succeed.

The key observation is that HOBET  is an expansion around momenta $k \sim 1/b$,
and thus differs from EFT approaches that expand around $k \sim 0$.   Consequently a
HOBET $P$-space lacks both high-momentum components important to short-range
$NN$ interactions and long-wavelength components important to minimizing the kinetic
energy.  While our group has previously discussed some of the consequences of
the combined infrared/ultraviolet problem in HOBET, here I identify another: a sharp
energy dependence in $H^{eff}$ that must be addressed before any simple representation
of $H^{eff}-H$ is possible.     This, combined with a trick to
remove operator mixing, leads to a simple and successful short-range expansion for
$H^{eff}-H$.
I conclude by noting how these results may set the stage for a successful determination
of the HOBET $H^{eff}$ directly from data.

\section{Review of the Bloch-Horowitz Equation}
The basis for the approach described here is the Bloch-Horowitz (BH) equation, which generates
a Hermitian, energy-dependent effective Hamiltonian, $H^{eff}$, which operates
in a finite Hilbert space from which high-energy HO Slater determinants are omitted:
\begin{eqnarray}
\label{wh:eq1}
H^{eff} = H &+& H {1 \over E - Q H} Q H \nonumber \\
H^{eff} |\Psi_P \rangle = E |\Psi_P \rangle ~~~&&~~~ |\Psi_P \rangle
 = (1-Q) |\Psi \rangle.
\end{eqnarray}
Here $H$ is the bare Hamiltonian and  $E$ and $\Psi$ are
the exact eigenvalue and wave function (that is, the results
of a full solution of the Schroedinger equation for $H$ in $P+Q$).  The BH equation
must be solved self-consistently, as $H^{eff}$ depends on $E$.  If this is done, the model-space calculation reproduces the exact $E$,
and the model-space wave function $\Psi_P$ is simply the restriction of $\Psi$ 
to $P$.  If one takes for $P$ a
complete set of HO Slater determinants with
HO energy $\le \Lambda_P \hbar \omega$, $H^{eff}$ will be 
translational invariant.
$P$ is then defined by two parameters, $\Lambda_P$ and the HO 
size parameter $b$.  
  
 The BH equation was solved numerically for the $av18$ potential using
 two numerical techniques.  In work carried out in
 collaboration with C.-L. Song \cite{song}, calculations were done for the deuteron and
 $^3$He/$^3$H by directly summing the effects of $av18$ in the $Q$-space.
 Because this potential has a rather hard core, sums to 140 $\hbar \omega$
 were required to achieve $\sim$ 1 keV 
 accuracy in the deuteron binding energy, and 70 $\hbar \omega$ to achieve
 $\sim$ 10 keV accuracies for $^3$He/$^3$H.  In work carried out with T. Luu \cite{luu}, such
 cutoffs were removed by doing momentum-space integrations over all possible 
 excitations.
 
 The results are helpful not only to the goals discussed previously, but also in illustrating 
 general properties of $H^{eff}$ that may not be widely appreciated.  For example,
 Table 1 gives the evolution of the $P$-space  $^3$He $av18$ wave function
 as a function of increasing $\Lambda_P$ for fixed $b$.
 $\Psi_P$ evolves simply, with each increment of
 $\Lambda_P$ adding new components to the wave function, while leaving previous
 components unchanged.  One sees that the probability of residing in the model space
 grows slowly from its 0$\hbar \omega$ value (31\%) toward unity. 
 
\begin{center}
\begin{table}
\tbl{Evolution of the $^3$He $av18$ HO wave function $\Psi_P$ with $\Lambda_P$ }
{\begin{tabular}{|r|r|r|r|r|r|r|} \toprule
 & \multicolumn{6}{c|}{amplitude} \\ \cline{2-7}
state & 0$\hbar \omega$ & 2$\hbar \omega$ &
4$\hbar \omega$ & 6$\hbar \omega$ & 8 $\hbar \omega$ & exact \\ \cline{2-7}
 & (31.1\%) & (57.4\%) & (70.0\%) & (79.8\%) & (85.5\%) & (100\%) \\ \hline
 $\mid 0, 1 \rangle$ & 0.5579 & 0.5579 & 0.5579 & 0.5579 & 0.5579 & 0.5579 \\ \hline
 $\mid 2, 1 \rangle$ & 0.0000 & 0.0463 & 0.0461 & 0.0462 & 0.0462 & 0.0463 \\ \hline
 $\mid 2, 2 \rangle$ & 0.0000 & -0.4825 & -0.4824 & -0.4824 & -0.4824 & -0.4826 \\ \hline
 $\mid 2, 3 \rangle$ & 0.0000 & 0.0073 & 0.0073 & 0.0073 & 0.0073 & 0.0073 \\ \hline
 $\mid 4, 1 \rangle$ & 0.0000 & 0.0000 & -0.0204 & -0.0204 & -0.0204 & -0.0205 \\ \hline
 $\mid 4, 2 \rangle$ & 0.0000 & 0.0000 & 0.1127 & 0.1127 & 0.1127 & 0.1129 \\ \hline
 $\mid 4, 3 \rangle$ & 0.0000 & 0.0000 & -0.0419 & -0.0420 & -0.0421 & -0.0423 \\ \hline
\end{tabular}}
\end{table}
\end{center}

Effective operators are defined by
\begin{equation}
\hat{O}^{eff} = (1 + HQ {1 \over E_f - HQ}) \hat{O}
(1 + {1 \over E_i - QH}QH)
\end{equation}
and must be evaluated between wave functions $\Psi_P$ having the nontrivial normalization
illustrated in Table 1 and determined by
\begin{equation}
1 = \langle \Psi| \Psi \rangle = \langle \Psi_P |
\hat{1}^{eff} | \Psi_P \rangle.
\end{equation}
The importance of this is illustrated in Fig. 1, where the elastic magnetic responses for
deuterium and $^3$He are first evaluate with exact wave functions $\Psi_P$ but bare
operators, then re-evaluated with the appropriate effective operators.   Bare operators
prove a disaster even at intermediate momentum transfers of 2-3 f$^{-1}$.  By using the effective
operator and effective wave function appropriate to $\Lambda_P$, the correct result --
the form factor is independent of the choice of $\Lambda_P$ (or $b$) -- is
obtained, as it must in any correct application of effective theory.

\begin{figure}
\begin{center}
\includegraphics[width=9cm]{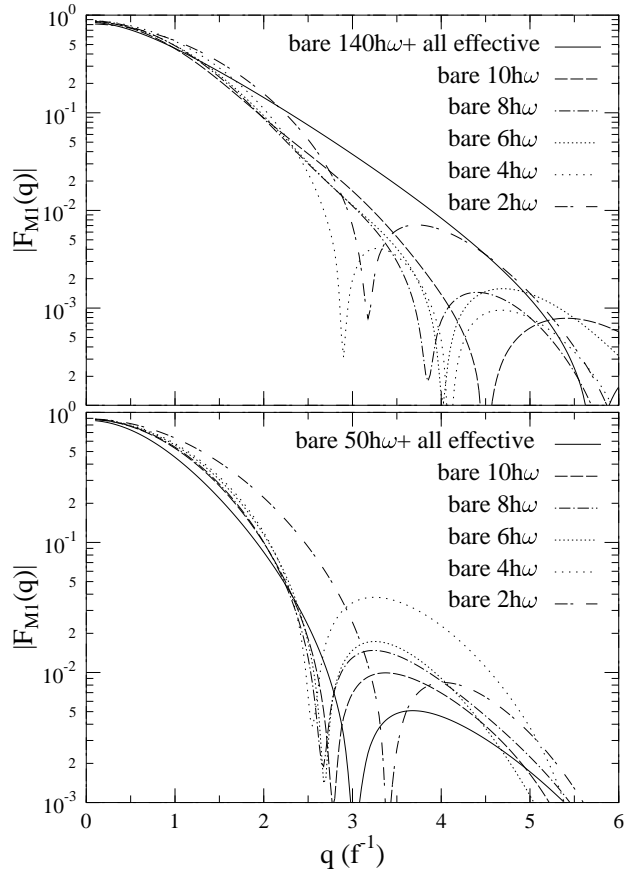}
\end{center}
\caption{Deuteron and $^3$He elastic magnetic form factors evaluated for various
$P$-spaces with bare operators (various dashed and dotted lines) and with the
appropriate effective operators (all results converge to the solid lines).}
\end{figure}

The choice of a HO basis excludes not only
high-momentum components of wave functions connected with the hard core,
but also low-momentum components connected with the proper asymptotic fall-off
of the tail of the wave functions.  This combined ultraviolet/infrared problem was first
explored by us in connection with the nonperturbative behavior of $H^{eff}$: the need 
to simultaneously correct for the missing long- and short-distance behavior of 
$\Psi_P$ is the reason one cannot define a simple $P$ that makes
evaluation of $H^{eff}$ converge rapidly.  We also found a solution to this problem,
a rewriting of the BH equation in which the relative kinetic energy operator is
summed to all orders.  This summation can be viewed as a transformation of a
subset of the Slater determinants in $P$ to incorporate the correct asymptotic falloff.
This soft physics, obtained from an infinite sum of high-energy HO states in $Q$,
is the key to making $H^{eff}$ perturbative.

It turns out this physics is also central to the issue under discussion here, the existence
of a simple representation for $H^{eff} = H + H { 1 \over E-QH} QH$, where $H=T+V$.
The reorganized BH $H^{eff}$ is the sum of the three left-hand-side (LHS) terms in Eqs. (\ref{wh:eq4})
\begin{eqnarray}
\langle \alpha | T + TQ {1 \over E-QT} QT | \beta \rangle&\underset{nonedge}{\longrightarrow}&\langle \alpha | T | \beta \rangle \nonumber \\
\langle \alpha | {E \over E-TQ} V {E \over E-QT} | \beta \rangle&\underset{nonedge}{\longrightarrow}&\langle \alpha | V | \beta \rangle \nonumber \\
\langle \alpha | {E \over E-TQ} V {1 \over E-QH} QV {E \over E-QT} | \beta \rangle&\underset{nonedge}{\longrightarrow}&\langle \alpha | V {1 \over E-QH} QV | \beta \rangle
\label{wh:eq4}
\end{eqnarray}

The first LHS term is the effective interaction for $T$, the relative kinetic energy.
As $QT$ acts as a ladder operator in the HO,
$E/E-QT$ is the identity except when it operates on an $|\alpha \rangle$
with energy $\Lambda_P\hbar \omega$  or $(\Lambda_P-1) \hbar \omega$.  We
will call these Slater determinants the edge states.   For nonedge states, this new expression and 
the BH form given in Eq. (\ref{wh:eq1}) both reduce to the expressions on the right of Eqs. (\ref{wh:eq4}).

Noting that the first LHS term in Eqs. (\ref{wh:eq4}) can be rewritten as
\begin{equation}
\langle \alpha | {E \over E-TQ} (T-{TQT \over E}) {E \over E-QT} | \beta \rangle,
\end{equation}
we see that the $QT$ summation can be viewed as a transformation to a new basis for $P$, 
$E/(E-QT) |\alpha \rangle$, that is orthogonal but not orthonormal.  This edge-state basis builds in
the proper asymptotic behavior governed by $QT$ (free propagation)
and the binding energy $E$.
The transformation preserves translational
invariance, as $T$ is the relative kinetic energy operator. Viewed in the transformed basis,
the appropriate effective interaction in given by the LHS terms in the square brackets in
Eqs. (\ref{wh:eq6}) below.

Alternatively, the results can be viewed as two equivalent expressions for the effective
interaction between HO states, but with a different division between ``bare" and
``rescattering" contributions
\begin{eqnarray}
\mathrm{bare:}{E \over E-TQ} \left[ H - {TQT \over E}  \right] {E \over E-QT}&\Leftrightarrow&H \nonumber \\
\mathrm{rescattering:}{E \over E-TQ} \left[ V {1 \over E-QH} QV \right] {E \over E-QT}&\Leftrightarrow& H {1 \over E-QH} QH
\label{wh:eq6}
\end{eqnarray}
{\it It is this new division that is critical.}  The expressions are identical for nonedge states.
But for edge states, only the expression on the left isolates a quantity, $VGV$, that is short-range
and nonperturbative.  We will see that this is the term that can be represented by
a simple, systematic expansion.  

Figure 2 shows the extended tail that is induced by $E/E-QT$ acting on a HO state.
Figure 3 is included to emphasize that there are important numerical differences between the two
expressions in Eqs. (\ref{wh:eq6}).  It compares calculations done for the deuteron using
the two ``bare" interactions:   thus in both cases $V$ enters
only linearly between low-momentum states, and all multiple scattering of $V$ in $Q$ in ignored.  
Figure 3 gives the resulting deuteron binding energy as a function of $b$, for several
values of $\Lambda_P$.   For the standard form
of the BH equation, a small model space overestimates the kinetic energy (too confining) and
overestimates short-range contributions to $V$ (too little freedom to create the needed
wave-function ``hole").  Making $b$ larger to lower the kinetic energy exacerbates the short-range
problem, and conversely.  Thus the best $b$ is a poor compromise that, even in
a 10 $\hbar \omega$ bare calculation, fails to bind the deuteron.  
But the new bare $H$ on the LHS of Eqs. (\ref{wh:eq6}) sums $QT$ to give
the correct wave-function behavior at large $r$, independent of $b$.
Then, for the choice $b \sim$ 0.4-0.5 f, the short-range physics can be absorbed directly into the
$P$ space.  The result is excellent 0th-order ground-state energy, with the residual
effects of multiple scattering through $QV$ being very small and
perturbative \cite{luu}.
 
\begin{figure}
\begin{center}
\includegraphics[width=10cm]{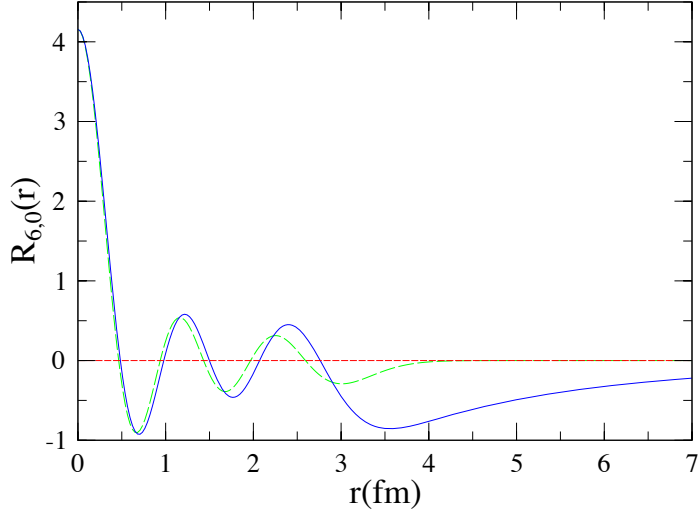}
\end{center}
\caption{A comparison of the $|n l\rangle$ and
extended $(E/E-QT) |n l\rangle$ radial wave functions, for the edge state $(n,l)=(6,0)$
in a $\Lambda_P=10$ deuteron calculation. Note that the normalization of the extended
state has been adjusted to match that of $|nl\rangle$ at $r$=0, in order to show that the
shapes differ only at large $r$.  Thus the depletion of the extended state at small $r$
is not apparent in this figure.}
\end{figure}

\begin{figure}
\begin{center}
\includegraphics[width=10cm]{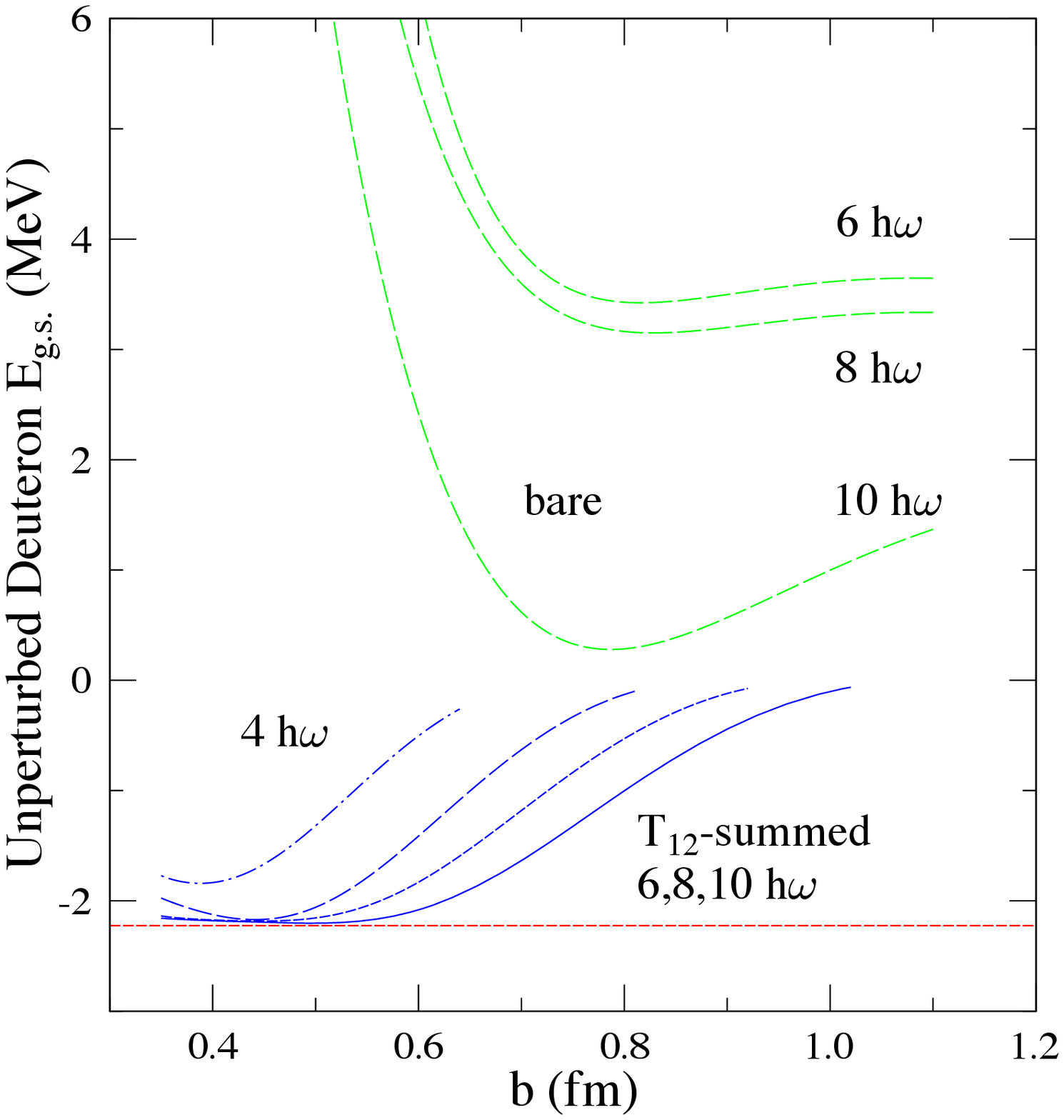}
\end{center}
\caption{Deuteron ground-state convergence in small $P$-spaces, omitting all effects
due to the multiple scattering of $V$ in $Q$.  The standard BH formulation with $P(T+V)P$ fails
to bind the deuteron, even with $\Lambda_P=10$.  The reorganized BH equation, where 
$QT$ has been summed to all orders but $V$ still appears only linearly, reproduces the correct binding energy for $\Lambda_P$=6.}
\end{figure}

\section{Harmonic-Oscillator-Based Effective Theory}
Now I turn to the question of whether (and how) the $Q$-space rescattering contribution
to $H^{eff}$  might be expressed
through some systematic short-range expansion.  There are two steps important in applying
such an expansion to HOBET.  One has to do with the form of the short-range expansion.
A contact-gradient (CG) expansion, constructed to include all possible
LO (leading order), NLO (next-to-leading order) , NNLO (next-to-next-to-leading order), ..., interactions
is commonly used,
\begin{eqnarray}
a_{LO}^{ss}(\Lambda_P,b) \delta({\bf r}) +
a_{NLO}^{ss}(\Lambda_P,b) (\overleftarrow{\nabla}^2 \delta({\bf r}) +
\delta({\bf r}) \overrightarrow{\nabla}^2)+ \nonumber \\
a_{NNLO}^{ss,22}(\Lambda_P,b) \overleftarrow{\nabla}^2 \delta({\bf r}) \overrightarrow{\nabla}^2 +
a_{NNLO}^{ss,40} (\Lambda_P,b)(\overleftarrow{\nabla}^4 \delta({\bf r}) + \delta({\bf r}) \overrightarrow{\nabla}^4).
\label{wh:eq5}
\end{eqnarray}

Because HOBET is an expansion around a typical momentum scale $\sim~1/b$, rather than
around $\vec{k}=0$, it is helpful to redefine the derivatives appearing in the CG expansion
Noting
\begin{equation}
\overrightarrow{\nabla}^n \exp{i \vec{k} \cdot \vec{r}}~\arrowvert_{k=0} = 0, n=1,2,...., 
\end{equation}
we demand by analogy in HOBET
\begin{equation}
\overrightarrow{\nabla}^n \psi_{1s}(b) = 0, n=1,2,... 
\end{equation}
This can be accomplished by redefining the operators $\hat{O}$ of Eqs. (\ref{wh:eq5}) by
\begin{equation}
\hat{O} \rightarrow e^{r^2/2} \hat{O} e^{r^2/2}
\end{equation}

The gradients in Eq. (\ref{wh:eq5}) then act on polynomials in $r$, a choice that removes 
all operator mixing.
That is, if $a_{LO}$ is fixed in LO to the $n=1$ to $n=1$ matrix element, where $n$ is
the nodal quantum number, it remains fixed in NLO, NNLO, etc.  Furthermore, the expansion
is in nodal quantum numbers, e.g., 
\begin{equation}
\overrightarrow{\nabla}^2 \sim (n-1)~~~~~~~~\overrightarrow{\nabla}^4 \sim (n-1)(n-2)
\end{equation}
so that matrix elements become trivial to evaluate in any order.  It can be shown that
the leading order in $n$ contribution agrees with the plane-wave result, and that
operator coefficients are a generalization of standard Talmi integrals for nonlocal potentials, e.g.,
\begin{equation}
 a_{NNLO}^{ss,22} \sim \int^\infty_0 \int^\infty_0 e^{-r_1^2} r_1^2 V(r_1,r_2) r_2^2 e^{-r_2^2} r_1^2 r_2^2 dr_1 dr_2
\end{equation}

The next step is to identify that quantity in the BH equation that should be identified
with the CG expansion.  This has to do with the two forms of the BH equation discussed
previously.  Consider the process
of progressively integrating out $Q$ in favor of the CG expansion, beginning
at $\Lambda >> \Lambda_P$
and progressing to $\Lambda=\Lambda_P$.  Using the projection operator
\begin{equation}
Q_\Lambda = \sum_{\alpha=\Lambda_P+1}^{\Lambda} | \alpha \rangle \langle \alpha |~~~~
Q_{\Lambda_P} \equiv 0
\end{equation}
we can isolate the contributions, above some scale $\Lambda$, to the two BH rescattering
terms of Eqs. (\ref{wh:eq6})
\begin{eqnarray}
\Delta(\Lambda)&=&H {1 \over E-QH} QH - H {1 \over E-Q_\Lambda H} Q_\Lambda H \nonumber \\
\Delta_{QT}(\Lambda)&=&{E \over E-TQ} \left[ V {1 \over E-QH} QV - V {1 \over E-Q_\Lambda H} Q_\Lambda V \right] {E \over E-QT}.
\end{eqnarray}
The goal of a CG expansion might be successful reproduction of the matrix elements of
$\Delta(\Lambda)$ -- the $Q$-space rescattering contributions for the standard form of the
BH equation -- as $\Lambda \rightarrow \Lambda_P$.  This
would allow us to replace all $Q$-space rescattering by a systematic short-range expansion, 
opening the door to a purely phenomenological determination of $H^{eff}$ for the SM.
The test case will be an  8$\hbar \omega$ $P$-space calculation for the deuteron
($\Lambda_{P}$ = 8, $b$=1.7 f).   The running of the 15 independent $^3$S$_1$ matrix elements of $\Delta(\Lambda)$
are plottted in Fig. 4a.  Five of  these are distinguished because they involve
an edge-state bra or ket (or both).  The evolution of these 
contributions with $\Lambda$ is seen to be somewhat less regular than that of
nonedge-state matrix elements.   The results for $\Lambda=\Lambda_P$ show that rescattering
is responsible for typically 12 MeV of binding energy.
 
The CG fit to the results in Fig. 4a were done in LO,  NLO, and NNLO as a function of $\Lambda$,
using the standard form of the BH equation.
The coefficients are fit to the lowest-energy matrix elements.
Thus in LO $a_{LO}(\Lambda)$
is fixed by the $1s - 1s$ matrix element, leaving 14 unconstrained matrix elements;
the NNLO fit ($1s-1s$, $1s-2s$, $1s-3s$, and $2s-2s$) leaves 11 matrix elements
unconstrained.  This is easily done, because the operators do not mix; e.g., among these
four, only the $1s-3s$ matrix element is influenced by
$a_{NNLO}^{ss,40}$.
The result is a set of coefficient that run as a function of 
$\Lambda$ in the usual way, with $a_{LO}$ small and dominant for large $\Lambda$, and
with the NLO and NNLO terms turning on as the scale is dropped.
Figs. 4b-d show the residuals -- the differences between the predicted and calculated matrix elements. 
For non-edge-state matrix elements the scale at which typical
residuals in $\Delta$ are significant, say greater than 100 keV (above $\sim$ 1\%), is brought down successively,
e.g., from $\sim 100 \hbar \omega$, to $\sim 55 \hbar \omega$ (LO), to $\sim 25 \hbar \omega$
(NLO), and finally to $\sim \Lambda_P \hbar \omega$ (NNLO).  But matrix elements involving edge
states break this pattern: the improvement is not significant, with noticeable deviations remaining at 
$\sim 100 \hbar \omega$ even at NNLO.  

\begin{figure}
\begin{center}
\includegraphics[width=12cm]{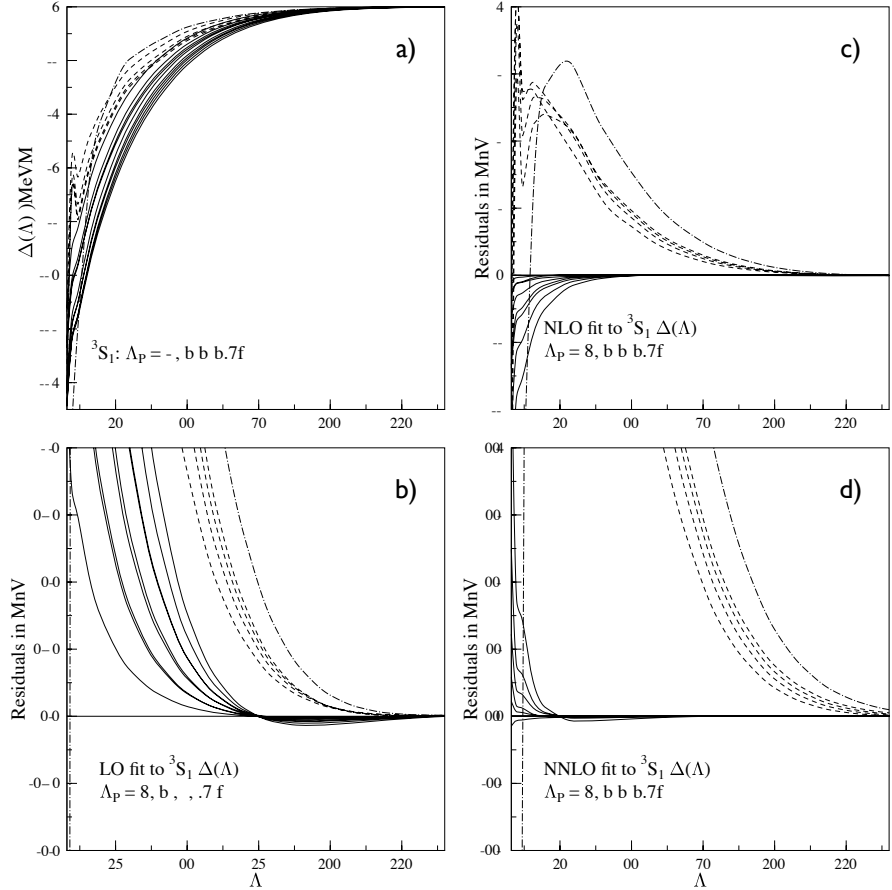}
\end{center}
\caption{In a) rescattering contributions to $H^{eff}$ due to excitations in $Q$ above $\Lambda$ 
are given for the standard form of the BH equation.   These results are for the 15 matrix elements
that arise in an 8$\hbar \omega$ calculation for the deuteron, with $b$=1.7 f.
In b)-d) the residuals of LO, NLO, and NNLO fits are shown (see text).  Matrix elements with
bra or ket (dashed) or both (dot-dashed) edge states are seen not to improve systematically.}
\end{figure}

This failure could be anticipated: because $QT$ strongly couples nearest shells across the
$P-Q$ boundary, $H{1 \over E-QH} QH$ contains long-range physics.  The candidate 
short-range interaction is $V {1 \over E-QH} QV$, not $H{1 \over E-QH} QH$: 
this is the reason $QT$ should be first summed,
putting the BH equation in a form -- the LHS of Eqs. (\ref{wh:eq6}) --
that isolates this quantity.  This reorganization affects edge-state matrix elements only,
those with the large residuals in Figs. 4b-d.  

To use the reorganized BH equation, the $QT$
sums appearing in Eqs. (\ref{wh:eq4}) must be completed.  There are several procedures for
doing this, but one convenient method exploits the raising/lowering properties of
$T$.  The result is a series of continued fractions $\tilde{g}_i(2 E/\hbar \omega, \{\alpha_i\},\{\beta_i\})$,
where $\alpha_i =(2n+2i+l-1/2)/2$ and $\beta_i = \sqrt{(n+i)(n+i+l+1/2)}/2$.  For any
operator $\hat{O}$ (e.g., $V$, $V {1 \over E-QH} QV$, etc.) 
\begin{equation}
\langle n' l' | {E \over E-TQ} O {E \over E-QT} | n l \rangle = \sum_{i,j=0} \tilde{g}_j(n',l') \tilde{g}_i(n,l) \langle n'+j~ l | O |n+i~ l \rangle
\end{equation}
It follows that the coefficients of the CG expansion for a HO basis must be
redefined for edge states, with a state- and E-dependent renormalization
\begin{eqnarray}
&&a_{LO} \rightarrow a_{LO}'(E/\hbar \omega ; n',l',n,l) = a_{LO}  \sum_{i,j=0} \tilde{g}_j(n',l') \tilde{g}_i(n,l) \nonumber \\
&&\times \left[ {\Gamma(n'+j+1/2)
\Gamma(n+i+1/2) \over \Gamma(n'+1/2) \Gamma(n+1/2)} \right]^{1/2}
\left[{(n'-1)! (n-1)! \over (n'+j-1)! (n+i-1)!} \right]^{1/2} .
\end{eqnarray}
This renormalization, which introduces no new parameters, can be evaluated in a similar way
for heavier systems: $T$ remains a raising operator.

\begin{figure}
\begin{center}
\includegraphics[width=12cm]{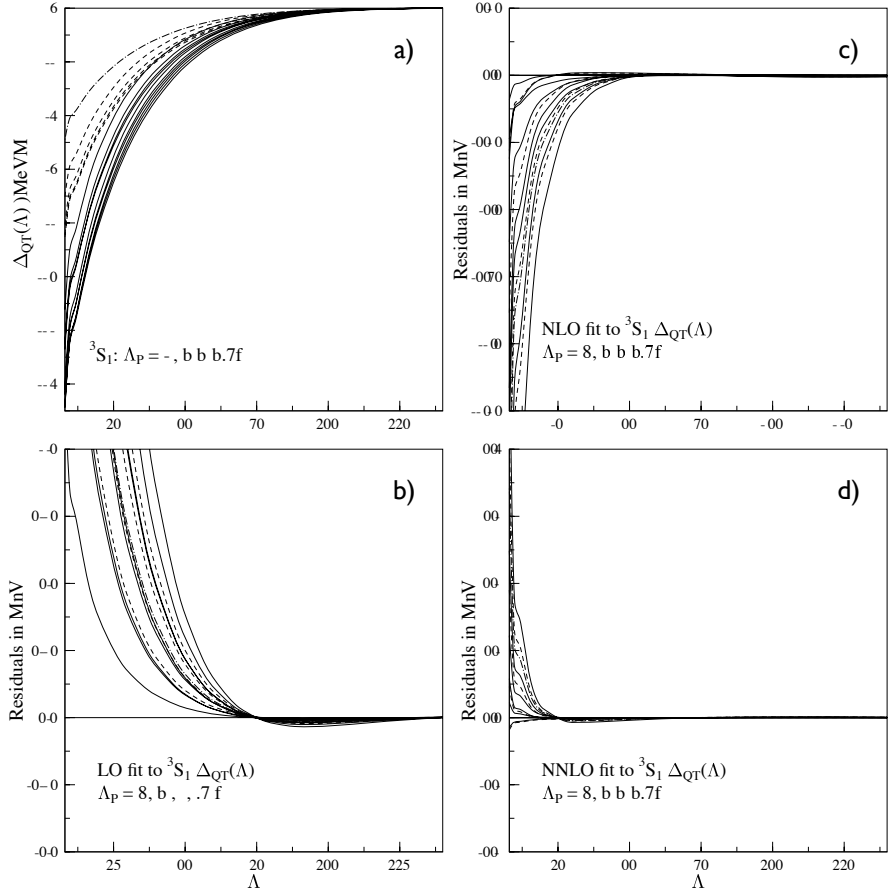}
\end{center}
\caption{As in Fig. 4, but with the edge states treated according to the $QT$-summed
reorganization of the BH equation, as described in the text.}
\end{figure}

Figs. 5 shows the results: the difficulties encountered for $\Delta(\Lambda)$ do not arise
for $\Delta_{QT}(\Lambda)$.  The edge-state matrix elements are
now well behaved, and the improvement from LO to NLO to NNLO is systematic
in all cases.
When $\Lambda \rightarrow \Lambda_P$, the CG potential continues to reproduce
$H^{eff}$ for the $av18$ potential
remarkably well, with an the average error in $^3$S$_1$
NNLO matrix elements of about 100 keV (or 1\% accuracy).  Other channels we explored 
behaved even better: the average error for the 15 $^1$S$_0$ matrix elements is about 10 keV (or 0.1\%
accuracy).   Because all matrix elements of $H^{eff}$ are reproduced well, the CG potential
preserves spectral
properties, not simply properties of the lowest energy states within $P$.
The NNLO calculation in the $^3$S$_1$-$^3$D$_1$ channel yields a deuteron ground-state energy
of -2.21 MeV.

Several points can be made:\\
$\bullet$  The net effect of the $QT$ summation is to
weaken the CG potential for HO edge states: the resulting, more extended state has
a reduced probability at small $r$.  Consequently the
effects of $QV$ are weaker than in states immune from the effects of $QT$. \\
$\bullet$  The very strong $QT$ coupling of the $P$ and $Q$ spaces is clearly problematic for
an ET: small changes in energy denominators alter the induced interactions.  Thus it is
quite reasonable that removal of this coupling leads to a strong energy dependence in the
effective interaction between HO states.  I believe that proper treatment of this energy
dependence will be crucial to a correct description of the bound-state spectrum in the HO
SM. \\
$\bullet$  This process can also be viewed as a transformation to a new, orthogonal 
(but not normalized) basis for $P$ in which $|\alpha \rangle \rightarrow {E \over E-QT} |\alpha \rangle$.
This yields basis states with the proper asymptotic behavior for
each channel.  A CG expansion with fixed coefficients can be used between these states,
following the reorganized BH equation of Eqs. (\ref{wh:eq4}).\\
$\bullet$  While our calculations have been limited to the deuteron, this same phenomena must
arise in heavier systems treated in HO bases -- $QT$ remains the ladder operator.  As the
issue is extended states that minimize the kinetic energy, it is clear that the relevant parameter must be the Jacobi coordinate associated with the lowest breakup channel.  This could be an issue for
treatments of $H^{eff}$ based on the Lee-Suzuki transformation, which transforms the
interaction into an energy-independent one .  In approaches like the no-core shell model \cite{nocore},
the Lee-Suzuki transformation is generally not evaluated exactly, but instead
only at the two-body level.  If such an approach were applied, for example, to
$^6$Li, a system weakly bound (1.475 MeV) to breaking up as $\alpha$+d, it is not obvious that 
a two-body Lee-Suzuki transformation would treat the relevant Jacobi coordinate responsible
for the dominant energy dependence.  This should be explored.\\
$\bullet$  I believe the conclusions about the CG expansion will apply to other effective
interactions.  For example, V-low-k \cite{achim}, a
soft potential obtained by integrating out high-momentum states, is derived in a plane-wave
basis, where $T$ is diagonal.  Thus it should be analogous to our CG
interaction, requiring a similar renormalization when embedded into a HO SM space.  It would
be interesting to test this conclusion.   

\section{Summary}
These results show that the effective interaction in the HO SM must have a very sharp dependence
on the binding energy, defined as the energy of the bound state relative to the first open channel.
This is typically 0 to 10 MeV for the bound states of most  nuclei (and 2.22 MeV for the 
deuteron ground state explored here).  Once
this energy dependence is identified, the set of effective interaction  
matrix elements can be represented quite well by a CG expansion,
and the results for successive LO, NLO,
and NNLO calculations improve systematically.

This result suggests that the explicit energy dependence of the BH equation is
almost entirely due to $QT$ --  though this inference, based on the behavior of matrix elements
between states with different number of HO quanta, must be tested in a case where multiple
bound states exist.

We also presented a simple redefinition of the gradients associated with CG
expansions, viewing the expansion as one around a momentum scale $\sim 1/b$.  This
definition removes operator mixing, making NNLO and higher-order fits very simple.  The expansion
then becomes one in nodal quantum numbers, with the coefficients of the expansion related
to Talmi integrals, generalized for nonlocal interactions.

While our exploration here has been based on ``data" obtained from an exact BH
calculation of the effective interaction for the $av18$ potential, this raises the question, is such
a potential necessary to the SM?   That is, now that the success of an NNLO description
of $Q$-space contributions is established, could one start with $PHP$ and determine the 
coefficients for such a potential 
directly from data, without knowledge
of matrix elements of $H$ outside of $P$?   I believe the answer is yes, even in cases (like
the deuteron) when insufficient information is available from bound states.
It turns out that the techniques described here can be
extended into the continuum, so that observables like phase shifts could be combined with
bound-state information to determine the coefficients of such an expansion.  An effort of this
sort is in progress.

I thank M. Savage for helpful discussions, and T. Luu and C.-L. Song for enjoyable collaborations. 
This work was supported by the U.S. Department of Energy Division of Nuclear Physics and
by DOE  SciDAC grant DE-FG02-00ER-41132.

\end{document}